\documentclass[a4paper]{article}

\usepackage{INTERSPEECH2021}
\usepackage{subfigure}
\usepackage{amsmath}
\usepackage{multirow}
\usepackage{tabularx,colortbl}
\usepackage{soul}
\newcommand{\ctext}[3][RGB]{%
  \begingroup
  \definecolor{hlcolor}{#1}{#2}\sethlcolor{hlcolor}%
  \hl{#3}%
  \endgroup
}

\newcommand{\bftab}{\fontseries{b}\selectfont}
\title{Estimating articulatory movements in speech production with transformer networks}
\name{Sathvik Udupa$^1$, Anwesha Roy$^1$, Abhayjeet Singh$^1$, Aravind Illa$^2$*\thanks{*This work was done prior to joining Amazon when the author was part of the SPIRE lab, Indian Institute of Science.}, Prasanta Kumar Ghosh$^1$}
\address{
  $^1$Electrical Engineering, Indian Institute of Science (IISc), Bangalore-560012, India\\
  $^2$Amazon Alexa, Bangalore, India
  }
\email{}

\begin{document}

\maketitle
\begin{abstract}
We estimate articulatory movements in speech production from different modalities - acoustics and phonemes. Acoustic-to-articulatory inversion (AAI) is a sequence-to-sequence task. On the other hand, phoneme to articulatory (PTA) motion estimation faces a key challenge in reliably aligning the text and the articulatory movements. To address this challenge, we explore the use of a transformer architecture - FastSpeech, with explicit duration modelling to learn hard alignments between the phonemes and articulatory movements. We also train a transformer model on AAI. We use correlation coefficient (CC) and root mean squared error (rMSE) to assess the estimation performance in comparison to existing methods on both tasks. We observe 154\%, 11.8\% \& 4.8\% relative improvement in CC with subject-dependent, pooled and fine-tuning strategies, respectively, for PTA estimation. Additionally, on the AAI task, we obtain 1.5\%, 3\% and 3.1\% relative gain in CC on the same setups compared to the state-of-the-art baseline. We further present the computational benefits of having transformer architecture as representation blocks.
\end{abstract}


\section{Introduction}
\ctext[RGB]{255,255,0}{In between neuro-motor planning and speech acoustics, articulatory movements plays a vital role as an intermediate representation} \cite{denes1993speech}\ctext[RGB]{255,255,0}{. Linguistic information is conveyed by the neuro-motor planning in the brain in the form of discrete abstract units. Vocal muscles are activated by this information sent via motor nerves. This results in different temporally overlapping gestures of the articulators like tongue, lips, velum, larynx etc }\cite{goldstein2003articulatory,neuro}\ctext[RGB]{255,255,0}{. These articulatory gestures in turn modulate the spectrum of acoustic signal which results in speech sound wave.}

Knowledge about the articulatory position along with the acoustics is useful in applications like automatic speech recognition \cite{ref8,ref9}, language learning \cite{S2018,8462401} and speech synthesis \cite{Illa2019,ling2009integrating}. In practice, in the absence of direct articulatory movements, they are typically estimated from different modalities like acoustic features (Mel-Frequency Cepstral Coefficients (MFCC)) and text (phoneme sequence). A rich literature exists related to the estimation of articulatory movements from acoustic features of speech, known as acoustic-to-articulatory inversion (AAI). Various approaches have been proposed, including Gaussian Mixture Model (GMM) \cite{ref12}, Hidden Markov Model (HMM) \cite{ref14}, and neural network \cite{ref13, wu2015acoustic}. The state-of-art performance is achieved by the bidirectional long short-term memory (BiLSTM) networks \cite{illa2018low,BLSTM}.

On the other hand, there have been few attempts on phoneme to articulatory (PTA) mapping even though it has many potential applications like in language tutoring systems for learning correct pronunciation or for analyzing pronunciation defects \cite{HMMP2A}. These attempts typically used the techniques from the speech synthesis paradigm like HMM \cite{HMMP2A} and BiLSTM \cite{BLSTMP2A}. Works in \cite{HMMP2A,BLSTMP2A} reported a considerable drop in phonological feature  (PTA model) performance compared to that using acoustic features (AAI model). One reason for the decline in performance using phonemes compared to acoustic features could be due to the limitations of the duration modelling with HMM \cite{HMMP2A,BLSTMP2A}. Hence, there is a need for effective duration modelling to obtain the articulatory frame length for a given phoneme sequence. In \cite{9053852}, the Tacotron \cite{taco2} speech synthesis  model is deployed for PTA task. Using the duration modelling in Tacotron, a soft alignment between the phonemes and the articulatory movement sequence is obtained. This is done by learning the attention weights, which implicitly model the time alignment between an encoder (phoneme representation) and decoder (articulatory representation) hidden states. However, due to the implicit duration modelling in Tacotron, it leads to poor convergence with a limited amount of training data \cite{9053852}. Further, sequential computations involved in the Tacotron network result in longer training and inference time. 

To overcome these limitations, this work explores the transformer networks for estimating articulatory movements. The transformer \cite{NIPS2017_3f5ee243} is a state-of-the-art sequence-to-sequence transduction model with an encoder-decoder structure. The architecture was first introduced on neural machine translation task, and recently, various transformer architectures \cite{articleTTS, ren2019fastspeech, lancucki2021fastpitch} have performed well in speech synthesis. We use a non-autoregressive transformer that exploits the entire sequence simultaneously. This allows for significantly more parallelization within training examples, which is important for longer sequences.  We deploy the architecture presented in FastSpeech \cite{ren2019fastspeech} for the PTA task. This allows for explicit duration modelling allowing hard alignment between the phonemes and the articulatory movement frames.

In addition to the PTA task, we also use a transformer model for the AAI task. To the best of our knowledge, this is the first work to utilize transformer networks for articulatory estimation. We perform extensive experiments with different positional encoding schemes and evaluate the results on different experimental setups. We compare the proposed approach with baseline models, and experimental results demonstrate that transformers achieve 154\%, 11.8\% and 4.8\% relative improvement in terms of correlation coefficient (CC) on subject-dependent, pooled and fine-tuned setups, on the PTA task, while the relative gains in CC are, respectively, 1.5\%, 3\% and 3.1\% on the AAI task.

A set of 460 phonetically balanced English sentences are considered from the MOCHA-TIMIT corpus as the stimuli for data collection from 10 subjects comprising of 6 male (M1, M2, M3, M4, M5, M6) and 4 female (F1, F2, F3, F4) subjects with ages in the range of 20-28 years. All the subjects are native Indians who are proficient in English and did not report any speech disorders in the past. All subjects were familiarized with the 460 sentences to prevent any elocution errors during recording. For each sentence, we simultaneously recorded audio signals using a microphone \cite{EM9600} and articulatory movement data using Electromagnetic Articulograph (EMA) AG501 \cite{AG501}. EMA AG501 has 24 channels to measure the horizontal, vertical, and lateral displacements as well as angular orientations of a maximum of 24 sensors. The sensors were placed according to the guidelines provided in [17] and the articulatory movement was recorded at a sampling rate of 250Hz.

Six sensors were glued on different articulators, viz. Upper Lip (UL), Lower Lip (LP), Jaw, Tongue Tip (TT), Tongue Body (TB), and Tongue Dorsum (TD). Additionally, two sensors were glued behind the ears for head movement correction. In this work, only the movements in the midsagittal plane are considered, corresponding to horizontal and vertical directions. Thus, we have twelve articulatory trajectories denoted by UL$_x$, UL$_y$, LL$_x$, LL$_y$, Jaw$_x$, Jaw$_y$, TT$_x$, TT$_y$, TB$_x$, TB$_y$, TD$_x$, TD$_y$. Manual annotations were done to remove silence at the start and end of each sentence.

\section{Proposed Approach}
We train different transformer models for AAI and PTA tasks. While an encoder-decoder structure is sufficient for AAI, additional duration modelling is required for PTA. In this section, we first summarise the transformer architecture and then describe the details involved in both tasks.  

\subsection{Transformers}   
As shown in Fig. \ref{speech_production_small}(a), the building block of a transformer layer consists of self-attention with residual connection and a 1D convolutional network.

\subsubsection{Self-attention}

We compute the self-attention as carried out in \cite{NIPS2017_3f5ee243}. \ctext[RGB]{255,255,0}{We construct three matrices query, key and value which are represented by \textit{Q}, \textit{K} and  \textit{V} respectively. As shown in Eqn. 1, these matrics are obtained by linear transformation of input $X$ through learnable weights,  \textit{$W^Q$},  \textit{$W^K$} and \textit{$W^V$} respectively. We then compute dot-product attention as shown in Eqn. 2, where $d$ is the feature dimension. } 
\begin{equation}
    Q = W^{Q}X, \hspace{4mm} K = W^{K}X, \hspace{4mm} V = W^{V}X  
\end{equation}

\begin{equation}
    Attention(\textit{Q}, \textit{K}, \textit{V} ) = 
    softmax(\frac{\textit{Q}\textit{K}^{T}}{\sqrt{d}})V 
\end{equation}

In multi-head attention, the self-attention function is calculated multiple times parallelly.  \ctext[RGB]{255,255,0}{Eqn. 3 represents multi head attention. The attentions obtained from different self attention blocks are concatenated together.} 

\begin{equation}
    MultiHead(Q, K, V ) = Concat(attn_{1},.., attn_{n})
\end{equation}
where,
\begin{equation}
    attn_m = Attention_m(Q_{m}, K_{m}, V_{m})
\end{equation}
This increases the model’s ability to focus on different positions in the sequence and it also gives the attention layer multiple different representation sub-spaces.     
\subsubsection{1D CNN}
Following the attention layer, there is a two-layer 1D convolutional network with ReLU activation. \ctext[RGB]{255,255,0}{Each sub-layer has a residual connection around it and is followed by layer normalization. That is, the output of each sub-layer can be written as LayerNorm($x$ + Sublayer$(x)$), where Sublayer$(x)$ is the function that the sub-layer itself implements. The residual connections help during training, by allowing gradients to flow through the networks directly. The layer normalizations stabilize the network which substantially reduces the training time necessary.}

Self-attention and 1D CNN form a transformer layer as shown in Fig. \ref{speech_production}(b). By stacking transformer layers sequentially, we obtain Feed Forward Transformer (FFT), which is used to represent encoder and decoder blocks.
    

\subsubsection{Positional Encoding}
\begin{figure}[htb]
   \centering
    \vspace{-.2cm}
   \centerline{\includegraphics[scale=0.51]{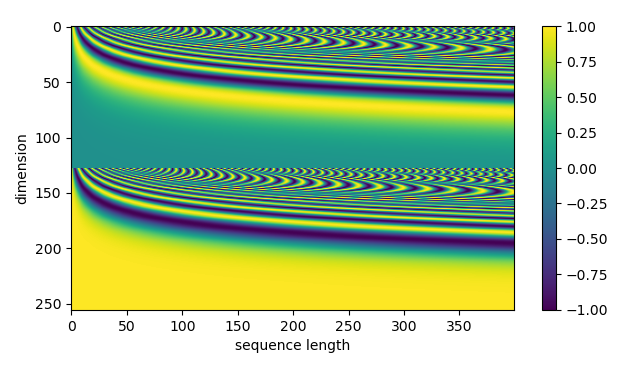}} 
   	\caption{Sinusoidal positional encoding}
   	\label{fig:SinuPos}
 \vspace{-.1cm}
\end{figure}

  Since there is no positional information encoded in the input sequence, positional encoding is combined with the input. This is usually done by adding or concatenating the sinusoidal positional encoding, as introduced in \cite{NIPS2017_3f5ee243}. \ctext[RGB]{255,255,0}{We have used a modified version used in implementation of }\cite{lancucki2021fastpitch}\ctext[RGB]{255,255,0}{, as shown in the equations below. It consists of sine and cosine functions concatenated together.}
  \begin{equation}
      PE(pos) = concat(sin(\omega*pos), cos(\omega*pos)) 
  \end{equation}
  where, 
 \begin{equation}
     \omega = \frac{1}{10000^{i/d}}
 \end{equation}
 \begin{equation}
     i = 0, 2, ..., (d/2)
 \end{equation}
  \ctext[RGB]{255,255,0}{The information present in the sinusoidal positional encoding in illustrated in Fig.}\ref{fig:SinuPos}. The advantage of using sinusoidal encodings is that they allow the model to extrapolate to longer sequence durations than the ones seen during training.
  
  Another way to introduce positional information is through relative positional embeddings as utilised in \cite{DBLP:journals/corr/abs-1803-02155}, \cite{DBLP:journals/corr/abs-1911-00203}. 
  
\begin{equation}
    e_{ij} = \frac{x_iW^Q(x_jW^K)^T+x_iW^Q({a_{ij}}^K)^T}{\sqrt{d}}
\end{equation}
where,
\begin{equation}
    a_{ij}^K = w_{clip(j-i, k)}^K
\end{equation}
\begin{equation}
    clip(x,k) = max(-k, min(k,x)) 
\end{equation}

\ctext[RGB]{255,255,0}{Eqn. 8 describes the self attention with relative postional encoding. It is the same attention mechanism illustrated in Eqn 2, with an additional term which contains the relative postional information. As shown in Eqn. 10, at every position $i$, positions ranging from $i-k$ to $i+k$ are awarded importance. Positional encodings are obtained for these positions from a learnable embedding, which acts as the extra term in self attention calculation.} While Sinusoidal encoding act on the entire sequence length, relative encoding allows for a small receptive field over which unique positional embeddings can be learnt.

\begin{figure}[htb]
  \centering
  \vspace{-.2cm}
   \centerline{\includegraphics[scale=0.35,]{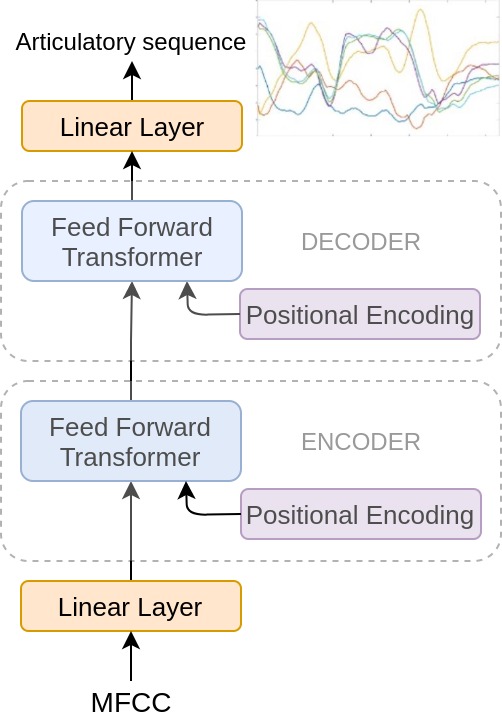}}
  
  \caption{Transformer arhitecture for AAI}\label{speech_production_aai}
  \vspace{-.4cm}
\end{figure}

\subsection{AAI with transformers} 
\ctext[RGB]{255,255,0}{The architecture for AAI with transformers is shown in Fig. }\ref{speech_production_aai}. We use 2 dense layers as an embedding, followed by an encoder-decoder structure consisting of FFT architecture. \ctext[RGB]{255,255,0}{The embedding is necessary to represent the input MFCCs in a higher dimension feature space as required by the transformer network. We also find that without 2 dense layers, the self-attention matrix in the first transformer layer learns vertical alignment, which makes the attention mechanism redundant. We expect the self-attention to learn dependecies along the sequence, which is well represented by a monotonic attention alignment}. Although position information is implicitly encoded as temporal information in acoustics, it may fail over modelling long term dependencies. Further, with MFCC features as input, adding sinusoidal encoding may harm the model performance as explored in previous work \cite{2018arXiv180309519S}. So we perform AAI experiments with different ways of inserting positional information. We first try with no positional encoding with the assumption that it is present implicitly in MFCCs. We then insert sinusoidal information by adding or concatenating it with the input. Finally, we use relative positional embedding with a total receptive field of 10 time-steps on either side of each position in the sequence.

\begin{figure}[htb]
  \centering
  \vspace{-.2cm}
   \centerline{\includegraphics[width=8cm]{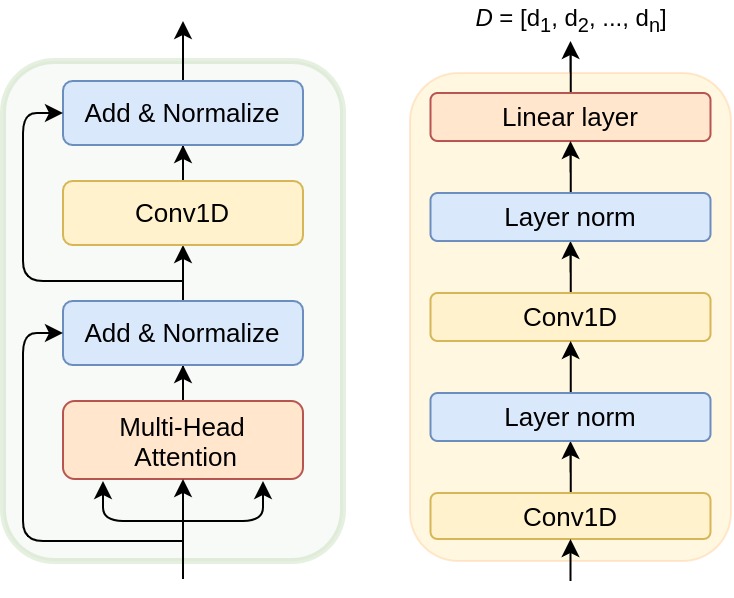}}
  
  \caption{(a) Transformer Layer (b) Duration predictor for PTA task }\label{speech_production_small}
  \vspace{-.4cm}
\end{figure}

\subsection{PTA with FastSpeech}
\begin{figure}[htb]
  \centering
  \vspace{-.2cm}
 \centerline{\includegraphics[scale=0.35]{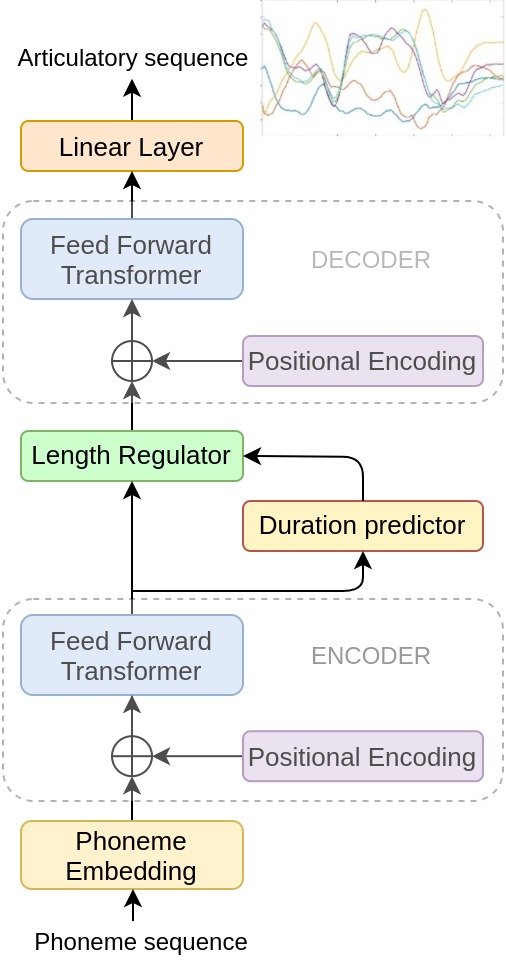}}
  
  \caption{FastSpeech architecture for PTA task }\label{speech_production}
  \vspace{-.4cm}
\end{figure}

We use the FastSpeech architecture as shown in Fig. \ref{speech_production} for the PTA task. \ctext[RGB]{255,255,0}{We first obtain a feature representation for the input phonemes indices using a learnable embedding. This acts as the input to the FFT Block which learns phoneme features. The output of the first FFT block is used to train the duration predictor.} A duration model is necessary for this task since there is no one-to-one correspondence between phonemes and articulator frames. \ctext[RGB]{255,255,0}{After obtaining the durations for each phoneme, the input to second FFT block is constructed from the phoneme features, using the length regulator.}

\subsubsection{Duration Predictor}

As shown in Fig. \ref{speech_production_small}(b), \ctext[RGB]{255,255,0}{the duration predictor consists of a two-layer 1D convolutional network with ReLU activation, where each layer is followed by layer normalization and dropout. This is followed by a linear layer which outputs a scalar value which is the predicted phoneme duration (length of articulatory sequence corresponding
to each phoneme).} This module is jointly trained with the main model with mean squared error as the loss function.

\subsubsection{Length Regulator}

Length regulator up-samples the phoneme sequence according to the predicted durations to obtain the length of articulator sequence.  For example, let the hidden states of a phoneme sequence be represented as $H_p = [h_1,h_2,h_3,h_4]$, with predicted phoneme durations $D = [2,2,3,1]$, then the expanded sequence becomes  $H_p = [h_1,h_1,h_2,h_2,h_3,h_3,h_3,h_4]$.

We use teacher forcing approach during training to obtain the decoder input length directly from ground truth durations. This is carried out to construct the decoder length accurately from the onset of training which helps in faster loss convergence. This also avoids unnecessary alignment or zero-padding to be performed during training.  

During inference, duration predictor outputs are used to construct the decoder input length. The phoneme encoder representation is replicated according to the number of frames to be present for each phoneme, which acts as the input to the decoder. At the output, we remove the zero-padded positions and perform DTW alignment as carried out in \cite{9053852} using the Euclidean distance metric between the predicted and ground-truth articulatory trajectories and then report the metrics.  \ctext[RGB]{255,255,0}{DTW alignment is necessary during inference in order to obtain evaluation metrics. Unlike the work done in }\cite{Biasutto-Lervat2018}\ctext[RGB]{255,255,0}{, the phoneme sequence has no timing information, which is why the lengths of articulatory movements have to be predicted. This is obtained in the form of individual phoneme durations. The predictions will not always match the ground truth, hence the need for DTW alignment.}

\subsection{Complexity Analysis}
 In transformers, the self-attention computation has $O(n^2 d)$ time complexity where $n$ is the sequence length and $d$ is the attention dimension. In LSTMs, it has $O(nd^2)$ complexity on every time-step and $O(n)$ complexity to visit the entire sequence. When $n$ is lesser $d$, a transformer layer processes faster than LSTM layer. On the PTA task, the phoneme sequence length at the encoder is 60, while the articulator length at the decoder is fixed at 400 for training. Due to the lower sequence length at the encoder, a large gain can be obtained on training and inference time. 
\section{Experimental Setup}
\subsection{Preprocessing}
The articulatory data is passed through a low-pass filter, with a cut-off frequency 25Hz, to remove high-frequency noise incurred during the measurement process. This also preserves the smoothly varying nature of the articulatory movement. We then perform sentence-wise mean removal and variance normalisation along each articulator dimension. The recorded speech is down-sampled from 48kHz to 16kHz, following which a forced alignment is performed to obtain phoneme sequence along with their boundaries using Kaldi \cite{povey2011kaldi} for every sentence. The resultant phonetic transcription of the dataset consists of 39 ARPABET symbols. We represent the phoneme input as a 1-dimensional vector with indices corresponding to the phonemes present in the sentence. We then zero-pad the vectors to obtain a fixed length equal to the maximum phoneme sequence length, which is set to 60. To represent acoustic data, we use 13-dimensional MFCC features. We zero-pad MFCCs and articulatory data to obtain a fixed length equal to the maximum number of frames, which is set to 400. 



\subsection{Model training and evaluation}
We perform both the PTA and AAI experiments on 10 subjects’ data with evaluation on unseen sentences. Each subject has 460 sentences from which 80\% of the sentences are used for training, 10\% for validation and 10\% for testing. We use root mean squared error (RMSE) and correlation coefficient (CC) \cite{doi:10.1121/1.3455847} as evaluation metrics to assess the performance of articulatory movement prediction. These are computed separately across each articulatory dimension and the average values across articulators and subjects are reported.

We examine the performance in 3 types of experimental setups namely subject-dependent (E1), pooled (E2) and fine-tuned (E3). In subject-dependent (E1) case, a model is trained for each subject and evaluated on that particular subject. In the pooled (E2) case, all subjects’ data are combined to train a single model and this is evaluated across all the subjects. In the case of fine-tuning (E3), we start with the pooled model as a pre-trained model and retrain for each subject separately and follow the evaluation as in E1. 

All experiments were performed using PyTorch \cite{NEURIPS2019_9015} on a single RTX 2080 GPU \ctext[RGB]{255,255,0}{with a batch size of 4. We use Adam optimizer with learning rate of $1e^{-4}$, along with a learning rate scheduler which reduces the learning rate by half every time the validation accuracy doesn't decrease in 7 epochs}. We use transformer architecture as implemented in NVIDIA’s open source code repository for FastPitch \cite{lancucki2021fastpitch}. \ctext[RGB]{255,255,0}{The codes for training our models, along with the hyperparameters used are available at https://github.com/bloodraven66/aai\_pta\_transformers.}  

\section{Results and Discussion}

We evaluate the performance of the proposed approach on AAI and PTA tasks using different experimental setups. We first present the results of the AAI task and then analyse the explicit duration modelling and predicted articulators from the PTA task. We also explore the benefits of having transformer models as representation blocks.

\subsection{\textbf{Performance on AAI}}
We train transformer models for AAI with different positional encoding, as shown in Table \ref{table:encodingResults}. We observe degradation in the performance using additive sinusoidal positional information. We obtain the best performance of 0.885 CC by using relative encoding in the case of E3, though there is a drop in performance in E1 using relative encoding compared to concatenative and no encoding models. We conclude that not using any encoding or using concatenative sinusoidal encoding works well generally when the amount of training data is not large. However, while relative encoding is useful with a large amount of data. 
\begin{table}[!htb]
\centering
\caption{AAI performance using the different models on various positional encodings. Standard deviation across all test sentences is reported in brackets}
\label{table:encodingResults}
\setlength{\extrarowheight}{2pt}
\resizebox{0.47\textwidth}{!}{\large
\begin{tabular}{|c|c|c|c|c|c|c|c|c|} 
\hline
\multirow{2}{*}{Setup} & \multicolumn{2}{c|}{None}                                                                                       & \multicolumn{2}{c|}{Additive}                                                                                   & \multicolumn{2}{c|}{Concatenative}                                                                              & \multicolumn{2}{c|}{Relative}                                                                                    \\ 
\cline{2-9}
                       & CC                                                     & RMSE                                                   & CC                                                     & RMSE                                                   & CC                                                     & RMSE                                                   & CC                                                     & RMSE                                                    \\ 
\hline
E1                     & \begin{tabular}[c]{@{}c@{}}0.854\\(0.03) \end{tabular} & \begin{tabular}[c]{@{}c@{}}1.062\\(0.08) \end{tabular} & \begin{tabular}[c]{@{}c@{}}0.839\\(0.03) \end{tabular} & \begin{tabular}[c]{@{}c@{}}1.115\\(0.09) \end{tabular} & \begin{tabular}[c]{@{}c@{}}\bftab0.857\\\bftab(0.02) \end{tabular} & \begin{tabular}[c]{@{}c@{}}1.060\\(0.08) \end{tabular} & \begin{tabular}[c]{@{}c@{}}0.836\\(0.02) \end{tabular} & \begin{tabular}[c]{@{}c@{}}1.351\\(0.11) \end{tabular}  \\ 
\hline
E2                     & \begin{tabular}[c]{@{}c@{}}0.871\\(0.02) \end{tabular} & \begin{tabular}[c]{@{}c@{}}1.017\\(0.08) \end{tabular} & \begin{tabular}[c]{@{}c@{}}0.867\\(0.02) \end{tabular} & \begin{tabular}[c]{@{}c@{}}1.038\\(0.08) \end{tabular} & \begin{tabular}[c]{@{}c@{}}0.874\\(0.02) \end{tabular} & \begin{tabular}[c]{@{}c@{}}1.013\\(0.08) \end{tabular} & \begin{tabular}[c]{@{}c@{}}\bftab0.884\\\bftab(0.02) \end{tabular} & \begin{tabular}[c]{@{}c@{}}0.971\\(0.08) \end{tabular}  \\ 
\hline
E3                     & \begin{tabular}[c]{@{}c@{}}0.877\\(0.02) \end{tabular} & \begin{tabular}[c]{@{}c@{}}0.999\\(0.09) \end{tabular} & \begin{tabular}[c]{@{}c@{}}0.874\\(0.02) \end{tabular} & \begin{tabular}[c]{@{}c@{}}1.018\\(0.09) \end{tabular} & \begin{tabular}[c]{@{}c@{}}0.881\\(0.02) \end{tabular} & \begin{tabular}[c]{@{}c@{}}0.996\\(0.08) \end{tabular} & \begin{tabular}[c]{@{}c@{}}\bftab0.885\\\bftab(0.02) \end{tabular} & \begin{tabular}[c]{@{}c@{}}0.968\\(0.09) \end{tabular}  \\
\hline
\end{tabular}}
\end{table}

In Table \ref{table:mainResults}, we further compare AAI results using BiLSTM and transformers with concatenative encoding. We observe 1.5\%, 3\% and 3.1\% relative improvement in CC using transformers over BiLSTM in subject-dependent (E1), pooled (E2) and fine-tuned (E3) setups respectively. We observe
that, although transformers have lesser parameters (Table \ref{table:timeMemoryUsage}), it performs better than BiLSTM on the AAI task.              

\subsection{\textbf{Performance on PTA}}
For the PTA task, we use the FastSpeech architecture with the duration predictor module. As shown in Table \ref{table:mainResults}, FastSpeech performs better than Tacotron in E1 setup, improving CC by 154\% (relative). This could mainly be due to the hard alignment obtained with the duration predictor allowing it to reliably predict articulator movements even with less data. On the other hand, the attention-based duration modelling in Tacotron needs more data to learn alignments. We apply the same duration predictor module on BiLSTM as encoder and decoder instead of transformers. However, it fails to learn good representation with BiLSTM as an encoder. Our experiments suggest that transformers are more compatible with the explicit duration predictor on the intermediate representation between the encoder and decoder.

\begin{table}[!htb]
\centering
\caption{PTA and AAI performance using different models in different training setups. Standard deviation across all test sentences is reported in brackets}
\label{table:mainResults}
\setlength{\extrarowheight}{2pt}
\resizebox{0.47\textwidth}{!}{\large
\begin{tabular}{|c|c|c|c|c|c|c|c|c|} 
\hline
\multirow{3}{*}{\large{Setup}} & \multicolumn{4}{c|}{PTA}                                                                                                                                                                                                                                                                                                     & \multicolumn{4}{c|}{AAI}                                                                                       \\ 
\cline{2-9}
                       & \multicolumn{2}{c|}{Tacotron   }                                                                                                                                                 & \multicolumn{2}{c|}{FastSpeech}                                                                                 & \multicolumn{2}{c|}{BiLSTM} 
                       & \multicolumn{2}{c|}{Tranformers}
                       \\ 
\cline{2-9}
                       & CC                                                    & RMSE                                                  & CC                                                    & RMSE                                                  & CC                                                    & RMSE                                                  & CC                                                    & RMSE                                                   \\ 
\hline
E1                     &   \begin{tabular}[c]{@{}c@{}}0.329\\(0.03)\end{tabular} & \begin{tabular}[c]{@{}c@{}}2.045\\(0.1)\end{tabular}  &
\begin{tabular}[c]{@{}c@{}}\bftab0.838\\\bftab(0.03)\end{tabular} & \begin{tabular}[c]{@{}c@{}}1.162\\(0.08)\end{tabular} &
\begin{tabular}[c]{@{}c@{}}0.844\\(0.02)\end{tabular} & \begin{tabular}[c]{@{}c@{}}1.05\\(0.08)\end{tabular} &
\begin{tabular}[c]{@{}c@{}}\bftab0.857\\\bftab(0.03)\end{tabular} & \begin{tabular}[c]{@{}c@{}}1.06\\(0.08)\end{tabular}
\\ 
\hline
E2                     &   \begin{tabular}[c]{@{}c@{}}0.682\\(0.04)\end{tabular} & \begin{tabular}[c]{@{}c@{}}1.482\\(0.09)\end{tabular} &
\begin{tabular}[c]{@{}c@{}}\bftab0.763\\\bftab(0.05)\end{tabular} & \begin{tabular}[c]{@{}c@{}}1.486\\(0.18)\end{tabular} &
\begin{tabular}[c]{@{}c@{}}0.849\\(0.02)\end{tabular}&
\begin{tabular}[c]{@{}c@{}}1.01\\(0.08)\end{tabular} &
\begin{tabular}[c]{@{}c@{}}\bftab0.875\\\bftab(0.02)\end{tabular} & \begin{tabular}[c]{@{}c@{}}1.013\\(0.08)\end{tabular}\\ 
\hline
E3                     &  \begin{tabular}[c]{@{}c@{}}0.806\\(0.03)\end{tabular} & \begin{tabular}[c]{@{}c@{}}1.18\\(0.08)\end{tabular}  &
\begin{tabular}[c]{@{}c@{}}\bftab0.845\\\bftab(0.03)\end{tabular} & \begin{tabular}[c]{@{}c@{}}1.124\\(0.08)\end{tabular} & 
\begin{tabular}[c]{@{}c@{}}0.854\\(0.02)\end{tabular} & \begin{tabular}[c]{@{}c@{}}0.99\\(0.085)\end{tabular}&
\begin{tabular}[c]{@{}c@{}}\bftab0.881\\\bftab(0.02)\end{tabular} & \begin{tabular}[c]{@{}c@{}}0.996\\(0.08)\end{tabular}\\
\hline
 \end{tabular}}
\end{table}
 In the E2 setup, we obtain 11.8\% improvement in CC over the Tacotron. We utilise these pooled models to fine-tune each subject. In E3, we observe an improvement of 4.8\% in CC. FastSpeech performs better than the Tacotron in all experimental setups although they have identical model size (Table \ref{table:timeMemoryUsage}).  

\begin{figure}[htb]
   \centering
   \centerline{\includegraphics[scale=0.5]{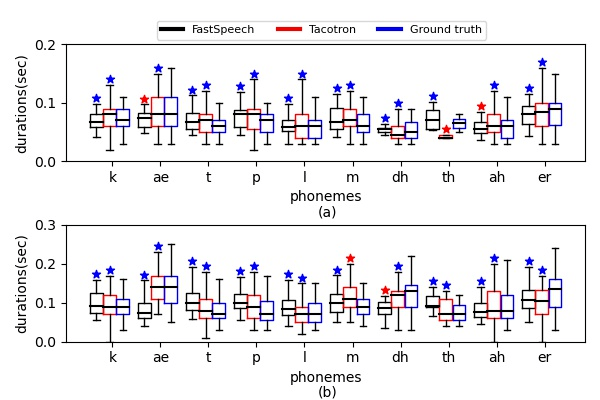}} 
   	\caption{Ground truth vs predicted durations for 10 phonemes for subjects (a) M1 and (b) F2. $\textcolor{blue}{\star}$ indicates the cases where the predicted duration is not statistically significantly (p$>$0.05) different from the ground truth duration. $\textcolor{red}{\star}$ indicates cases where they are significantly different (p$<$0.05)}
   	\label{fig:DurPlot}
\end{figure}

In Fig. \ref{fig:DurPlot}, we analyse the predicted durations from FastSpeech and Tacotron models in E3 setup on 10 phonemes. It is clear that the predicted duration in the case of subject M1 (Fig. \ref{fig:DurPlot}(a)) are not significantly (p$>$0.05) different than ground truth duration except for /ae/ \& /ah/ for FastSpeech and /th/ for Tacotron. With subject F2 (Fig. \ref{fig:DurPlot}(b)), a significant difference (p$<$0.05) between the ground truth and predicted duration is observed only in the case of /dh/ and /m/ for FastSpeech and Tacotron respectively. We observe that ground truth and predicted durations matches in the majority of the cases using Tacotron and FastSpeech on fine-tuned (E3) setup.

\begin{figure}[htb]
   \centering
   \centerline{\includegraphics[scale=0.5]{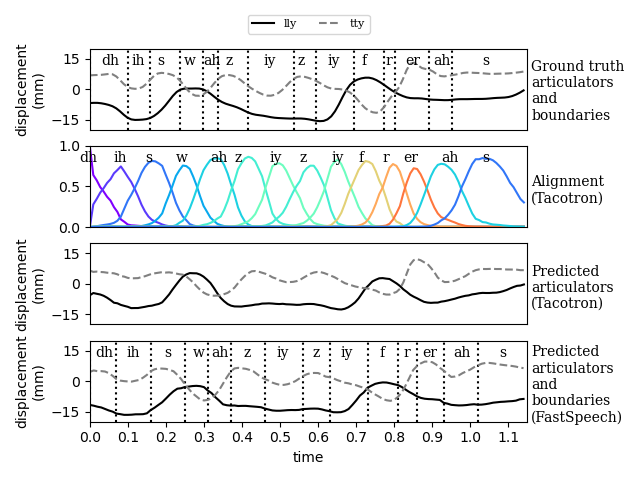}} 
    \vspace{-.4cm}
   	\caption{Comparing original LL$_y$ and TT$_y$ with the estimated ones from phonemes}
   	\label{fig:artPlot}
\end{figure}

In Fig. \ref{fig:artPlot}, we compare LL$_y$ and TT$_y$ ground truth articulatory trajectories along with FastSpeech and Tacotron predictions. On the ground truth articulatory movements, we mark phoneme boundaries obtained by forced alignment. We extract the predicted durations from FastSpeech and show it with the predicted articulators. For Tacotron, we use the attention alignment distribution from the trained network. 
We observe that FastSpeech predictions closely resemble the ground truth, both in phoneme durations and the articulatory trajectories. For example, consider $t=0.1$ when phoneme /ih/ is uttered, a dip is observed in LL$_y$ curve in ground truth and the FastSpeech predictions while it is flatter in the Tacotron prediction.        
\subsection{\textbf{Time-memory usage}}
We further compare the time and memory usage of the different models in Table \ref{table:timeMemoryUsage}. On the AAI task, we use BiLSTM with 3.7M parameters and a transformer Encoder-Decoder with 1.67M parameters. We observe transformer and BiLSTM have similar training time in AAI task. For the PTA task, we fix the number of model parameters at 27M for FastSpeech and Tacotron. We notice that FastSpeech is 6 times faster in E1 setup and 37 times faster in E2 setup training. This is possible due to significant parallelization on how representation is learnt across a sequence. We also observe a lot of speedup in inference time with FastSpeech.  From these observations, we can conclude that transformer architectures are better suited for PTA task. On comparing the transformer models on the two tasks, we notice that AAI uses 40 times smaller model compared to the PTA task, further work is required to bring down the model size of PTA models.

\begin{table}[!htb]
\centering
\caption{Time and memory usage of different models}
\label{table:timeMemoryUsage}
\setlength{\extrarowheight}{5pt}
\resizebox{0.47\textwidth}{!}{\large
\begin{tabular}{|c|c|c|c|c|} 
\hline
\multirow{2}{*}{Setup}                 & \multicolumn{2}{c|}{PTA} & \multicolumn{2}{c|}{AAI}  \\ 
\cline{2-5}
                                       & FastSpeech & Tacotron                     & Tranformers & BiLSTM      \\ 
\hline
E1 train time                          & 12 mins    & 74 mins                      & 7 mins     & 3 mins      \\ 
\hline
E2 train time                          & 21 mins    & 13 hrs                       & 45 mins      & 1 hr        \\ 
\hline
Inference time                         & 0.007 sec  & 0.44 sec                     & 0.003 sec   & 0.0006 sec  \\ 
\hline
\multicolumn{1}{|l|}{No of parameters} & 27M        & 27M                          & 1.67M       & 3.71M       \\
\hline
\end{tabular}}
\vspace{-0.5cm}
\end{table}


\section{Conclusions}

In this work, we propose the benefit of transformer based networks for the estimation of articulatory movements. We show relative improvement of  154\%, 11.8\% and 4.8\% in CC on subject-dependent (E1), pooled (E2) and fine-tuned (E3) setups respectively on the PTA task. We also show that having an explicit duration predictor is beneficial towards obtaining hard alignment with the articulator movements. This works particularly well on limited data scenarios. With transformer models for AAI, we improve the CC by 1.5\%, 3\% and 3.1\% on the same setups. We conclude that transformer network is a good alternative to learn articulator movements from phonemes and acoustics. We also observe that articulatory movements are predicted more accurately from acoustics (AAI) compared to phonemes (PTA). This is consistent with the findings from the past. In future, we plan on improving PTA to attain performance on par with AAI.  


\bibliographystyle{IEEEtran}

\bibliography{mybib}


\end{document}